# Spectroscopic Analysis of the Double Lined Eclipsing Binary αVir

M. I . Nouh[1,2] , S. M. Saad[2], I. Zaid[1], M. M. Elkhateeb[1,2] and E. Elkholy[1,2]

1- National Research Institute of Astronomy and Geophysics,  11421 Helwan, Cairo, Egypt.
2- College of Science, Northern Border University, 1320 Arar, Saudi Arabia.

**Abstract.**  αVir is a well known double-lined spectroscopic binary with a B-type for both components. In the present paper we have analyzed a total of 90 spectra obtained through 1992-2000. Spectral analysis are based on two spectral lines Hα and HeI 6678 belong to Hα region. Radial velocity analysis have suggested low eccentric orbit *(e=0.002)* at inclination (*i*=81°.89±2°.34) and with $4^d.01422$ period, semi-amplitude k1=*100 kms$^{-1}$,* mass ratio *q=0.49±0.05* and mass function *f(m)* =0.547. Using KOREL program for spectrum disentangling we have been able to decompose the spectrum of the system to its primary and secondary components. Synthetic spectral analysis of both the individual and disentangled spectra has been performed and yielded effective temperatures Teff= 25000±250 K, surface gravities log g=3.75±*0.25* and projected rotational velocities (v sin i= 180±5 km. s$^{-1}$) for the primary, while for the secondary they are Teff = 17000±250 K, log g = 4.0±0.25.

## 1. Introduction

αVir (Spica), (67Vir, HR5056, HD 116658, BD-l0° 3672, HIP 65474) is one of the brightest stars in the southern hemisphere (m$_v$ = 0.97). It is well known double-line spectroscopic binary (SB2) with short period (4.014 d), Shobbrook et al. (1969), and well resolved components, Popper (1980). Both components are B-type star (B1.5+B4), Popper (1980). The system are resolved using the intensity interferometry observations of Herbison-Evans et al. (1971) . Hipparcos Parallax is 12.44±0.86 *mas*.

The fundamental parameters of the components and the differences in their helium abundance are reported by Lyubimkov et al. (1995) based on high resolution and large S/N ratio spectra. Dukes (1974) from 722 photographic spectra has found four distinct periodic radial velocity variations between 4.2 - 6.6 hours superimposed on the primary orbital velocity and this has suggested further evidence that the primary is a β Cephei type star. Lomb (1978) indicated that, the pulsation has become unstable, and the amplitude has decreased to an undetectable level since 1972.

In the present paper, we analyze 90 spectra of the spectroscopic binary system $\alpha$ Vir, obtained through the years 1992-2000. Orbital solution for the system and disentangled spectra for the components are presented. Effective temperature, surface gravity and rotational velocities for both components are computed using model atmosphere anlaysis.

The structure of the paper is as follows. Section 2 is devoted to the observation and reduction of the spectra. Section 3 deals with the radial velocity measurements and orbital solutions. In section 4, the spectrum disentangling is outlined. Discussion of the results is represented in section 5.

## 2. Observations and Data Reduction

The available data for this study consisted of 90 spectroscopic observations in the range 6300-6700 Å and one file for the total spectra. The spectra were secured from Reticon archive of the Ondrejov 2-m Telescope. The spectra were obtained between 1992 and 2000. All initial reduction of spectra (i.e. wavelength calibration, zero level subtraction, flat-field correction and rectification) were carried out using the program SPEFO written by the late Dr. Jiřĭ. Horn, and described by Horn & Koubsky (1992), and also Skoda (1996).

## 3. RV Measurements and Orbital Solutions

Inspection of the present observed spectral range 6300-6700 Å, has shown that the principal feature is the presence of alternative shifting of two spectral lines, which insure the double-lined binarity of αVir. Fig; 1 shows the movement of the secondary profile in Hα and HeI 6678 lines as a consequence of the orbital motion.

### 3.1. RV Measurements

All the resulting fits files from the original sources have been analyzed using the program SPEFO. All acquisition times transformed to the heliocentric frame, and the measured radial velocities RVs are shifted to the zero-point using a set of telluric lines and the technique is described by Horn et al. (1996). Direct RVs measurements are obtained interactively by means of the best fit of the direct and reverse images of the measured line profile by using program SPEFO. The RVs are obtained from Hα and HeI 6678Å for the primary component. The RVs of the secondary component not obtained here, they are very weak and blended with those of the primary component through most of the orbital phases. Due to the continues variability of the line profiles we mainly have measured the most outer part of the lines. The measured RVs described above show radial-velocity variations with time, with semi amplitude of ∼100 *Km/s,* Fig 2 shows the measured RVs of the primary component from spectral lines Hα and HeI 6678, we can recognize that the semi-amplitude of RVs of HeI 6678 is slightly higher than that obtained for Hα.

## 3.2. Orbital Solution

Orbital elements were computed using FOTEL code Hadrava (1990) and the measured RVs of Hα and HeI 6678 lines. The initial spectroscopic elements were obtained in advance from SPEL program written by Dr. J. Horn. Two possible orbits were derived using FOTEL program, first we performed the orbital element for circular orbit, and then we obtained another solution for eccentric orbit. Through the circular solution, various values of the inclination **i** and mass ratio q were adopted in different runs. The best fit (with the minimum squared residuals) was obtained at **i** = $81^\circ.89 \pm 2^\circ.34$ and $q = 0.49 \pm 0.05$. Through the eccentric solution we fixed the mass ratio and the inclination angle to that values finally accepted for the circular solution and we adopted the rest of elements as free parameters then we iterated till we get the best fit of each value.

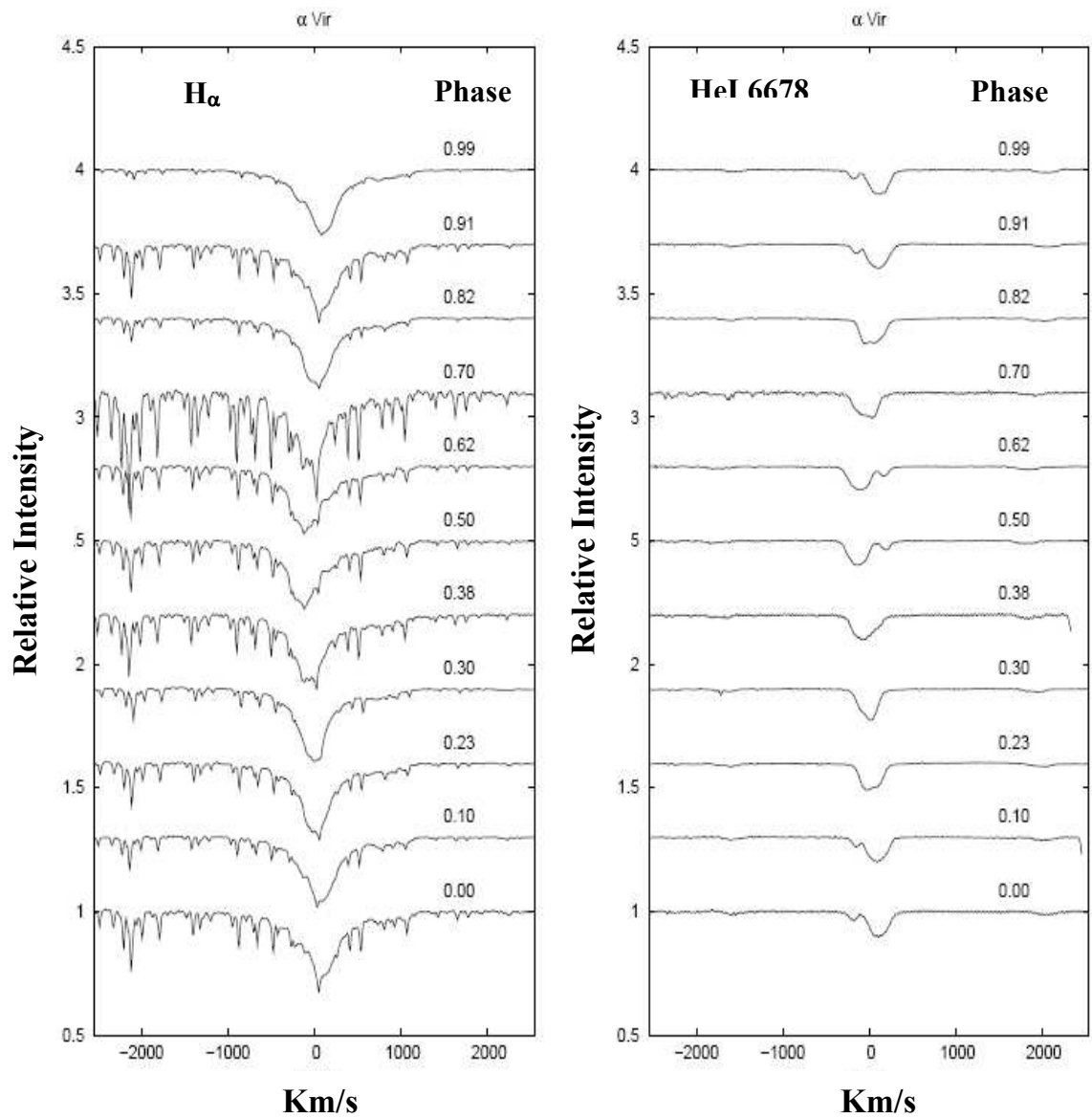

Fig 1: Evolution of Hα and HeI observed profiles of αVir. The profiles show the movement of the secondary according to the orbital motion. All the spectra are arbitrary shifted vertically to allow direct comparison.

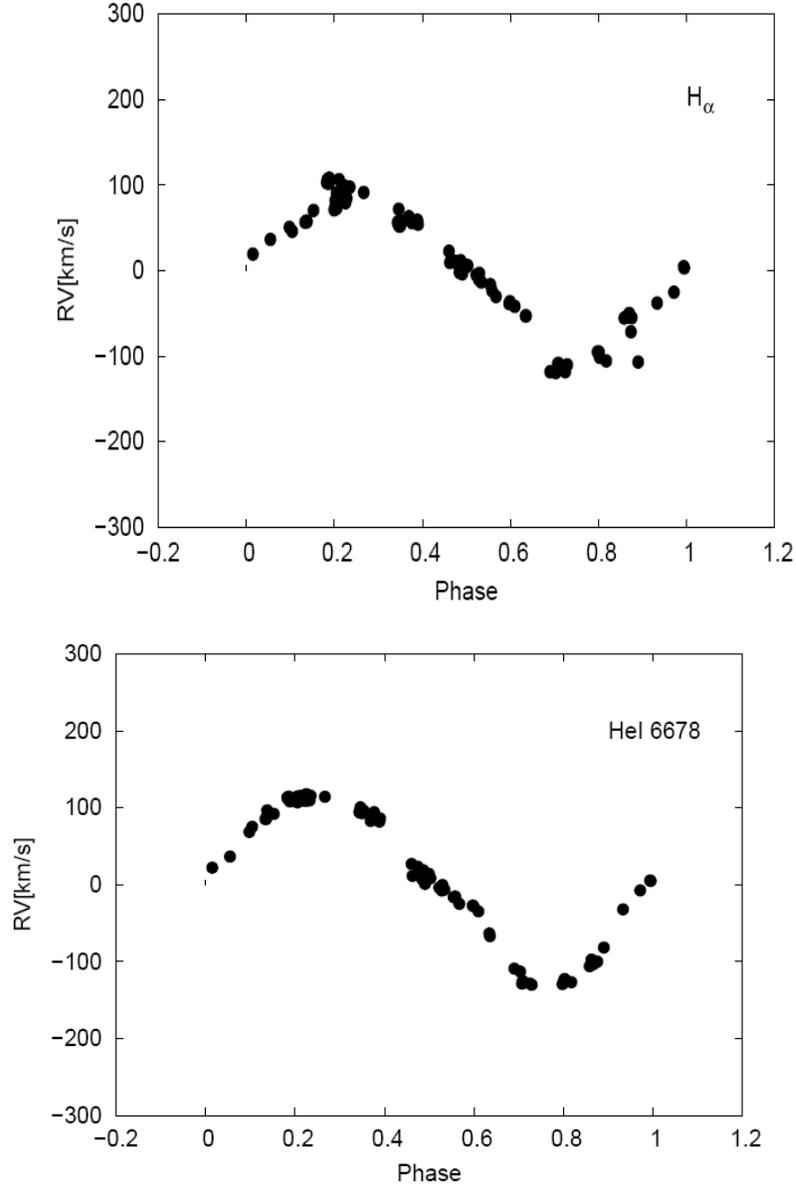

Fig. 2: Measured radial velocity curves for the primary of αVir. (Top panel): for Hα
(Lower panel): for HeI 6678, both are folded with the orbital period.

Orbital elements from both solutions are given in Table 1, Figures (1) and (2) represent the two ephemeris (corresponding to the time of maximum RVs) obtained from the circular and the eccentric solutions respectively. Figure (3) represents the results from FOTEL eccentric solution II, O-C residual from the computed solution and the corresponding velocity curve.

$$T_{max.RV} = (HJD2449499.77 \pm 0.022) + (4.^d01428 \pm 0.^d00092)\, X_E, \qquad (1)$$

and

$$T_{max.RV} = (HJD2449498.96 \pm 0.022) + (4.^d01422 \pm 0.^d00098)\, X_E. \qquad (2)$$

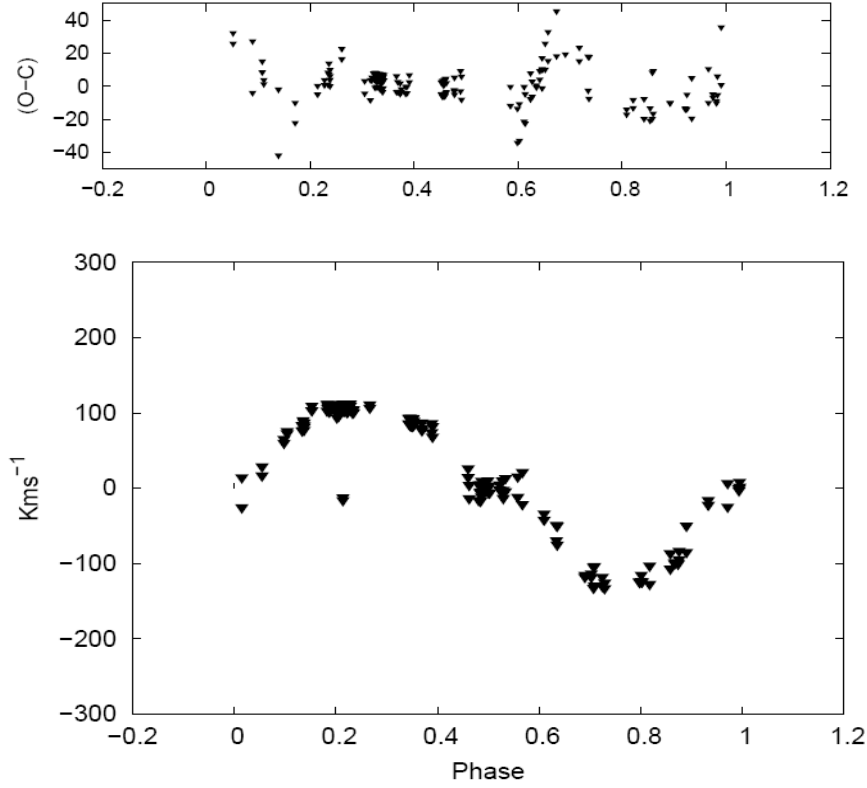

Fig. 3: (Upper panel): The residual (O-C), (lower panel): radial velocity curve based on FOTEL solution for the primary component of αVir, both are folded with the orbital period.

## 4. Spectrum Disentangling

Spectral disentangling is a method to derive individual spectra of the components of binary or multiple systems. The intensity ratios of the component stars are assumed to be the same at different phases, Hadrava (2004). In the present work two different spectral regions have been analyzed by means of spectral disentangling technique KOREL for Fourier disentangling developed by Hadrava (1997) has been used to produce separate spectra of αVir in the spectral regions 6510 – 6610 Å° and 6620 – 6740 Å° in simultaneous with the orbital elements computation. For the first region where many water lines are found, the decomposition of the telluric lines is also included. The binary system αVir is successfully resolved for its two components in both regions.

To proceed with KOREL we have adjusted both the parameters of circular and eccentric orbits. The eccentric solution is finally accepted for KOREL. KOREL orbital solutions from Hα and HeI 6678 regions are summarized in Table (1) solution III. Figs 4 and 5 (upper panels) represent the resulting decomposed line profiles for Hα and HeI 6678 of the primary and the secondary components, (lower panels) illustrate the corresponding velocity curves obtained from *KO*REL solution. Disentangling profiles from both regions are fairly revealed evidence of a secondary spectrum.

**TABLE 1: ORBITAL SOLUTION FOR αVir.**

| Element | FOTEL Sol. I<br>Hα+HeI 6678<br>Circular | FOTEL Sol. II<br>Hα+HeI 6678<br>Elliptical | KOREL Sol. III<br>Hα– HeI 6678 |
|---|---|---|---|
| $P[d]$ | 4.01428±0.00092 | 4.01422±0.00098 | 4.01426± 0.00076 |
| T *periast.* | 49499.77±0.022 | 49498.96±0.022 | 49498.76± 0.024 |
| $i$ (deg) | 81.89±2.34 | Fixed | |
| $k1$(km/s) | 109.61±1.51 | 109.56±15.9 | 95.02± 95.04 |
| $q(m2/m1)$ | 0.49±0.05 | Fixed | 0.476 ± 0.455 |
| E | 0.0 | 0.002±0.001 | 0.007± 0.003 |
| $\omega$ [deg] | 0.0 | 5.15±0.92 | 4.988± 0.82 |
| $\gamma_1$(km/s) | -3.82 | -3.86 | |
| $\gamma_2$(km/s) | -16.55 | -16.59 | |
| $f1(m)$ | 0.548E+00 | 0.547 | |
| $f2(m)$ | 0.466E+01 | 4.65 | |
| $M_1 sin^3 i$ | 0.103E+02 | 10.3 | |
| $M_2 sin^3 i$ | 0.507E+01 | 5.06 | |
| $M1(M\odot)$ | 0.106E+02 | 10.6 | |
| $M2(M\odot)$ | 0.522E+01 | 5.21 | |
| No. of RVs | 180 | 180 | |
| $\sum(O-C)^2$<br>$\sum(O-C)^2$ | 16.93<br>20.75 | 16.98<br>20.79 | |
| $\sum(O-C)^2$(Mean) | 18.94 | 18.98 | |

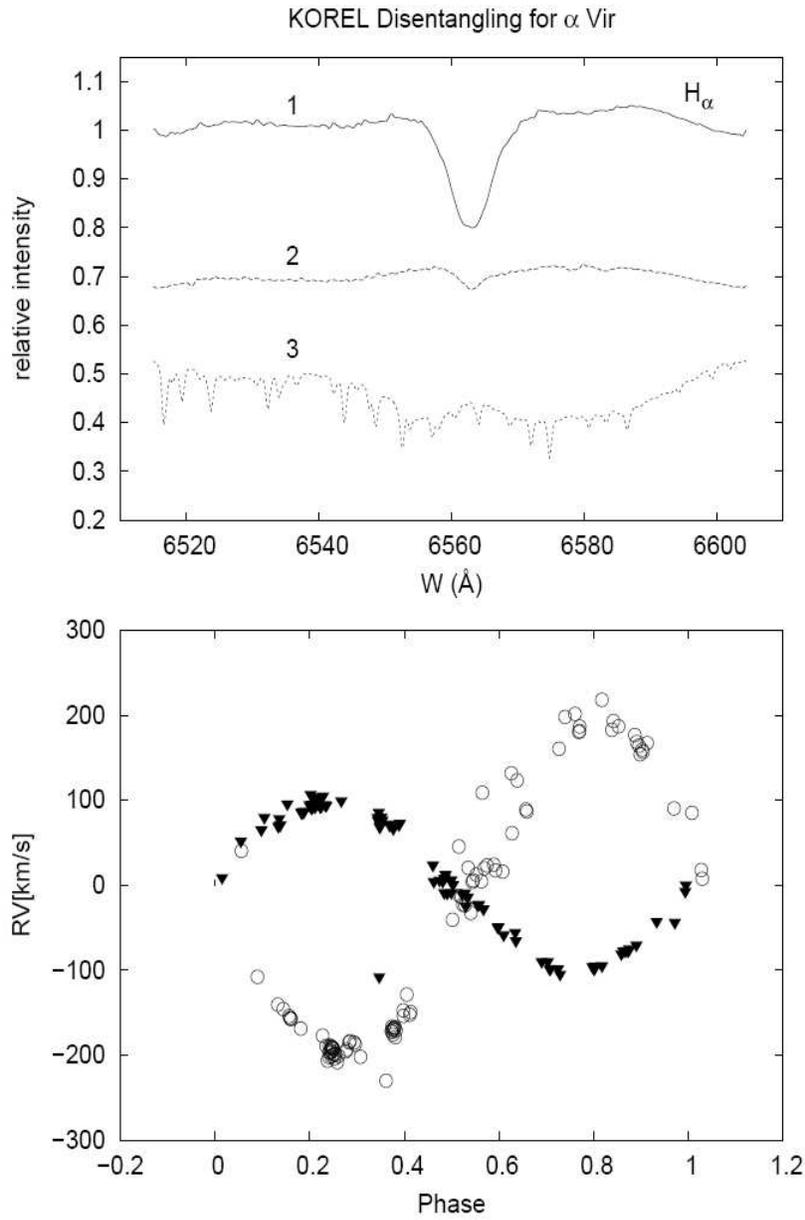

Fig. 4: Upper panel: The decomposed line Profiles of Hα line by KOREL. The numbers 1, 2 and 3 indicate profiles of the primary, the secondary and the telluric lines. The profiles are arbitrary shifted. Lower panel: the Hα radial velocity curve for both components as determined by KOREL. Filled triangles and open circules represent the primary and secondary respectively.

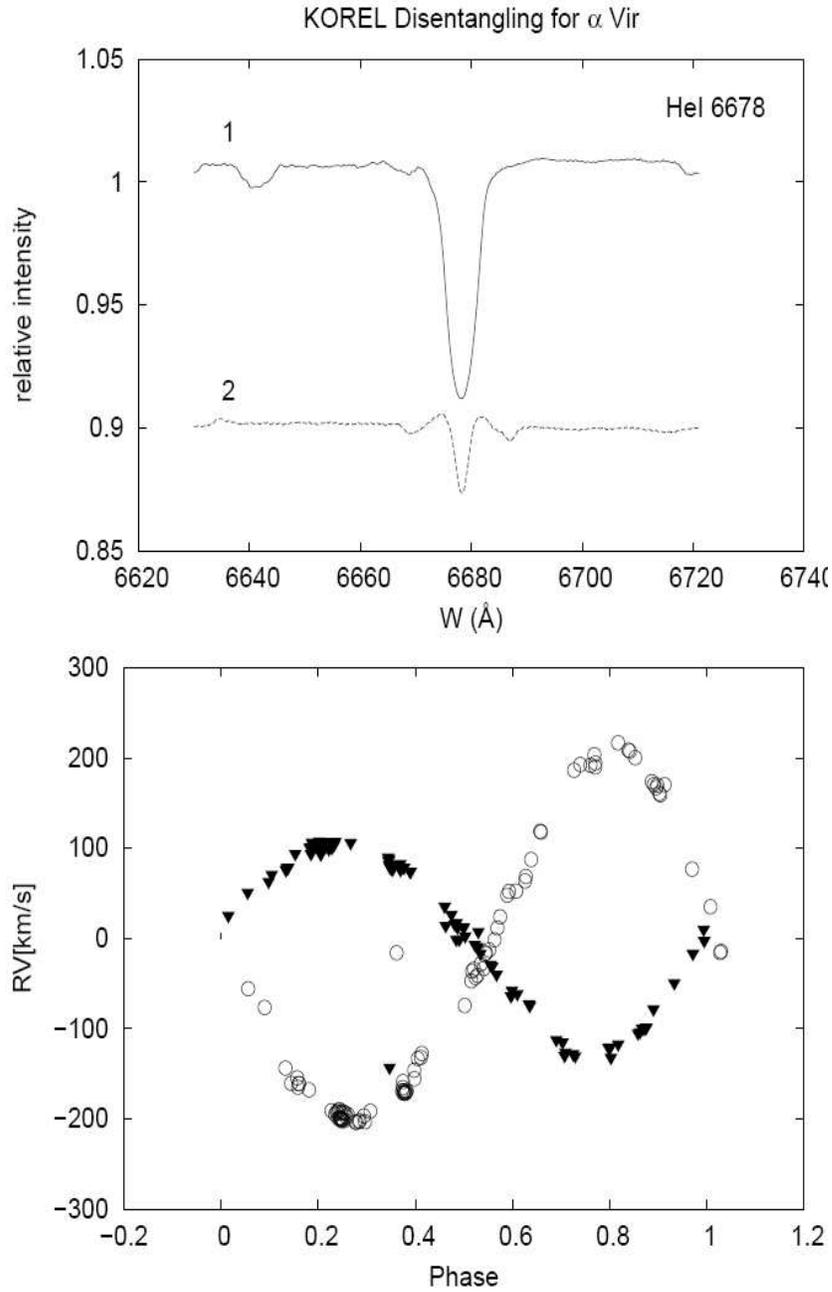

Fig. 5: Similar to the Fig. 4, for HeI 6678.

## 5. Model Atmosphere Analysis

We have adopted the NLTE grid of model spectra of early B type stars calculated by Hubeny and Lanz (2007). The grid is computed for the temperature range $15000 \leq T_{eff} \leq 30000$ K, with step 1000 K, surface gravity $1.75 \leq \log g \leq 4.75$ with step 0.25 dex. We adopt the grid computed at turbulent velocity $v_t = 2$ km/sec.

For $T_{eff} \leq 15000$ K, we have computed small grid of spectra using the program SPECTRUM (Gray 1992), the effective temperature takes the values from 10000 K to 12000 K with step 2500 K and two models with $T_{eff} = 13000, 14000$ K, the surface gravity takes the range $2.5 \leq \log g \leq 5$ with step 0.5 dex. We adopt as input model atmosphere that of ATLAS9, Kurucz (1993). All synthetic spectra are convolved (using the code ROTIN3) with Gaussian function having FWHM=0.25 to reduce the resolution of the synthetic spectra to the observed one.

We first used $\chi^2$ routine to compare the observed spectra with the theoretical one. The best fitted model is $T_{eff} = 25000 \pm 1000$ and $\log g = 3.75 \pm 0.25$. Using the code ROTIN3 we calculated a grid of rotated spectra for $5 \leq v \sin i \leq 500$ km/sec with step 5 km/sec. Then we applied $\chi^2$ again to determine the best rotational velocity for the two stars. To do this, we used the spectral line Mg II 4481 Å (the recommended line to be free from the pressure broadening, Gray (1976) in the fitting process. The comparison leads to rotational velocity $v \sin i = 180 \pm 5$ km/sec. The best fitted rotated model as well as the observed one is displayed in Figure () for the two lines He I 4471 Å and Mg II 4481 Å.

The observed and synthetic spectra are plotted in Figure (6), the wavelength range correspond to wavelength regions covered by the blue and red channel of the HEROS. As it is shown, there is a good agreement between the observed and model spectrum for most lines expect the cores of some Balmer lines.

To disentangle the effective temperature and surface gravity of the secondary, we used the total energy flux of the system which is related to the energy flux of the components A and B located at a distance d(pc) from the earth taking into account the radii of the two stars, Saad and Nouh (2011). The best fitted model gives, $T_{eff} = 17000 \pm 1000$ K and $\log g = 4 \pm 0.25$. Figure (7) shows the plot of the observed spectra at phase 0 (which almost represent the light came from the primary) along with the best fitted model spectra.

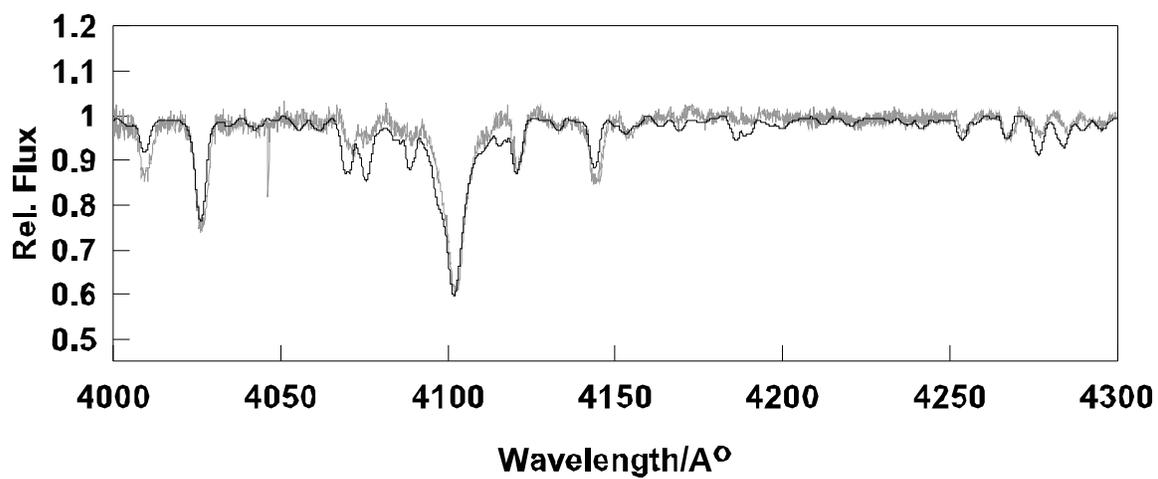
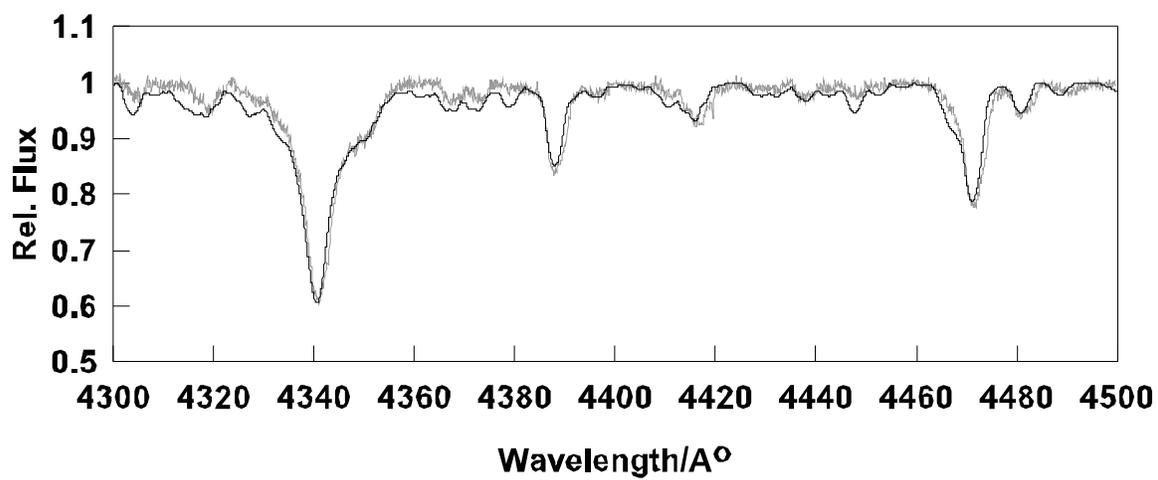

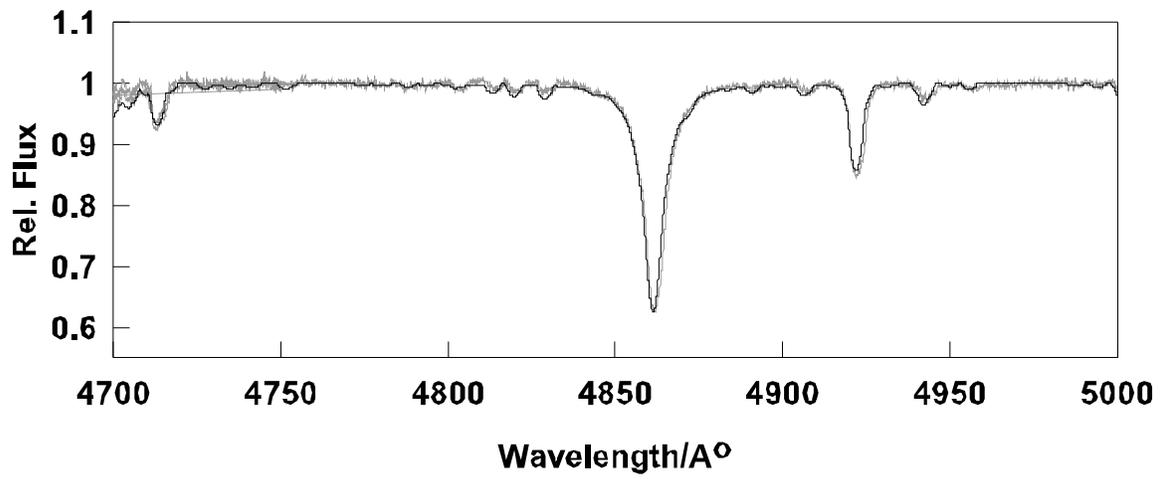

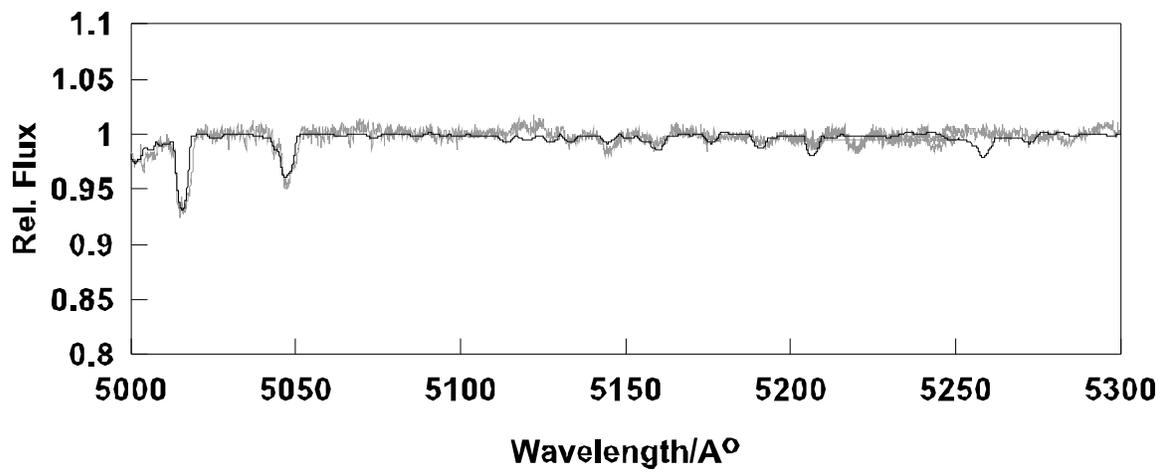

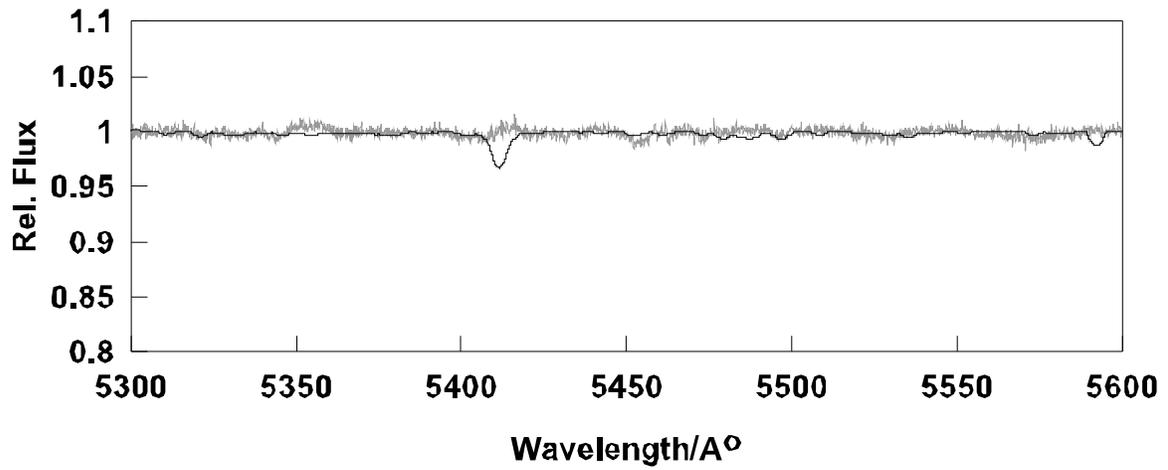

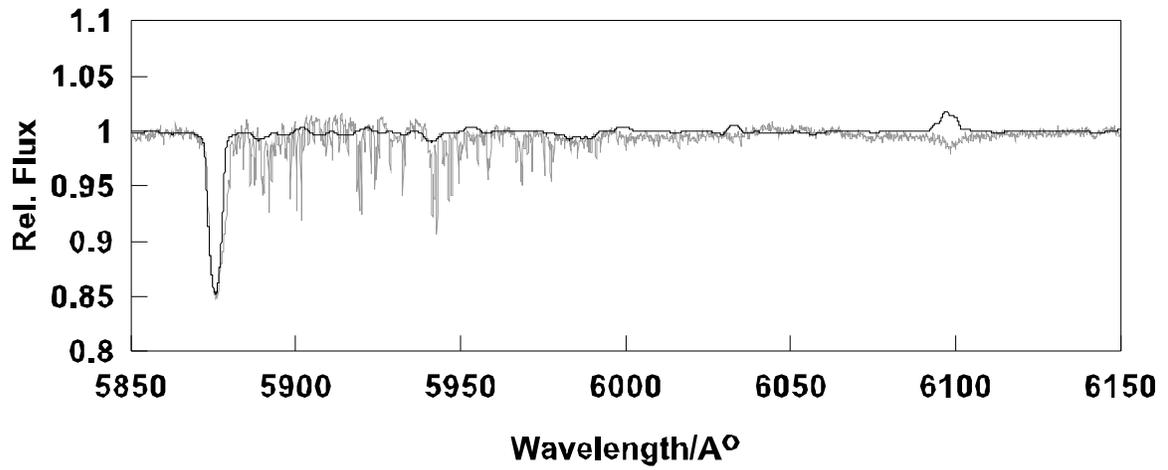

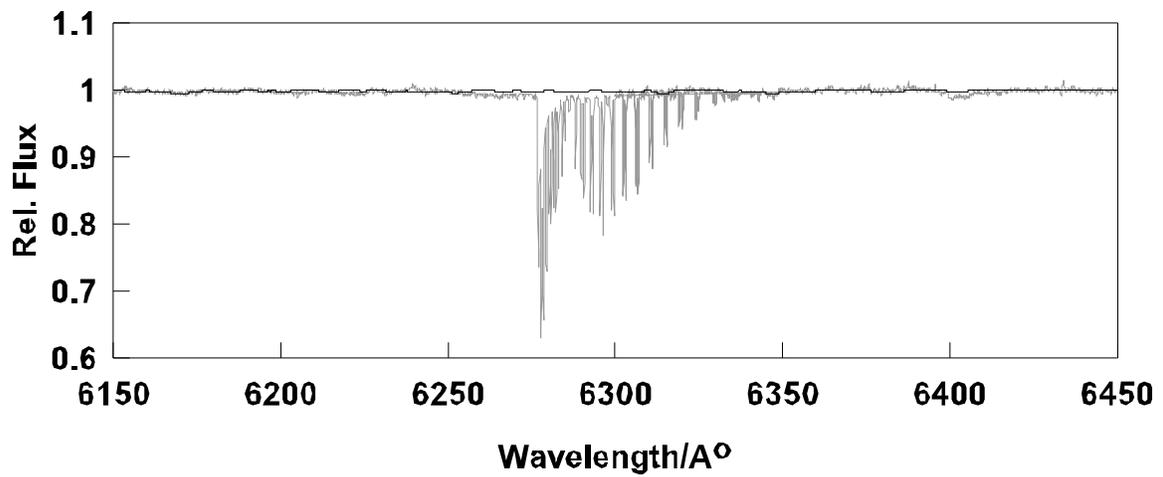

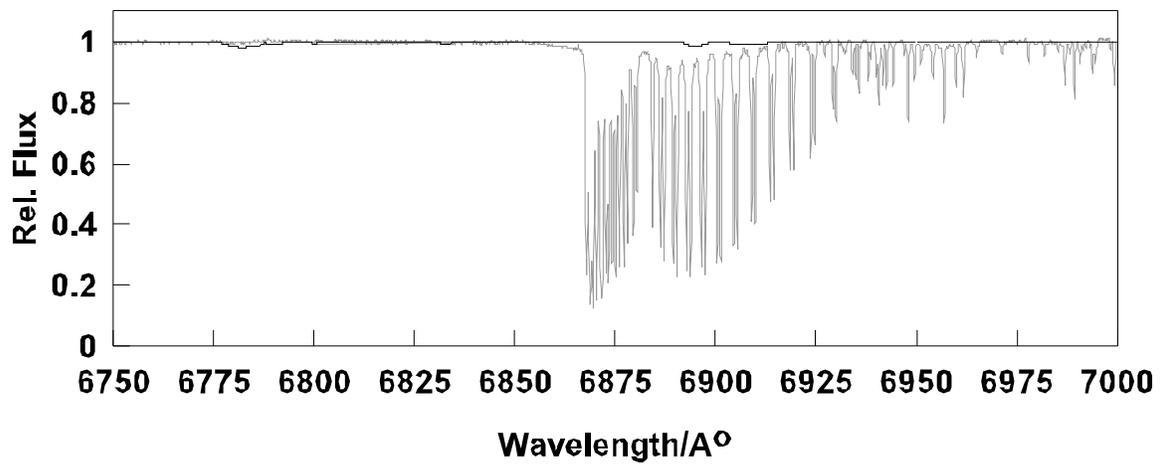

Figure (6). Line identification and comparison of the spectrum of α Vir (grey line) and the synthetic NLTE spectrum Teff = 25000K, Log g = 3.75 and v sin i = 180 Km/s (black line).

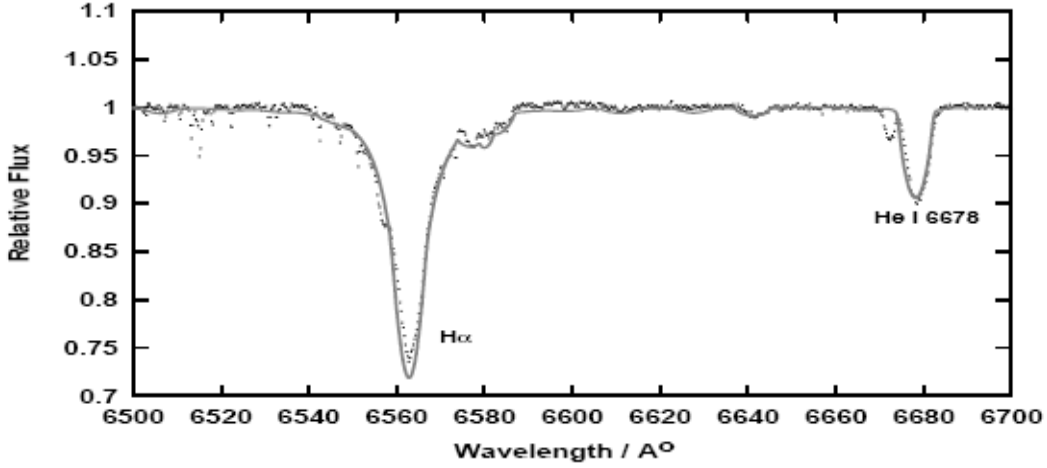

**Figure (7).** Comparison of observed spectrum of α Vir at phase 0 (dotted line) and the synthetic NLTE spectrum (grey line) $T_{eff}$ = 25000 K, log g =3.75, and v sin i = 180km· s$^{-1}$.

## 7. CONCLUSION

In summarizing the paper, the analysis of the new spectroscopic material has obtained for the SB2 αVir during the last decade allowed us to decompose the spectra to its individual components, to derive the orbital solution and to discuss the basic properties of the system. Orbital solutions from FOTEL and KOREL are based on the RVs of the primary. The accepted orbital solution suggests very low eccentric orbit at inclination 81°.89 and semi-amplitude ∼100 kms$^{-1}$. The solution also gives *q=0.49* and masses of M$_1$= 10.6 M$_\odot$ and M$_2$ =5.2 M$_\odot$.

Using KOREL program for spectrum disentangling we have been able to decompose the spectrum of the system to its primary and secondary components. Synthetic spectral analysis of both the individual and disentangled spectra has been performed and yielded effective temperatures Teff= 25000±250 K, surface gravities log g=3.75±*0.25* and projected rotational velocities (v sin i= 180±5 km. s$^{-1}$) for the primary, while for the secondary they are Teff = 17000±250 K, log g = 4.0±0.25.

**Acknowledgements:** This research has made use of the NASA's Astrophysics Data System Abstract Service and Ondřejov observatory 2-meter spectral archive. We would like to thank Prof. Dr. Hadrava (Author's of KOREL and FOTEL codes) for his helpfully discussions when using KOREL and FOTEL codes and for using his codes for free.